\documentclass[reprint,floatfix]{revtex4-1}
\usepackage{epsfig}
\usepackage{amsthm,amsmath}
\usepackage{color}

\begin{document}

\title{Calculation of P,T-odd interaction constant of PbF using Z-vector method in the relativistic coupled-cluster framework}
\author{Sudip Sasmal$^1$}
\email[]{sudipsasmal.chem@gmail.com}
\author{Himadri Pathak$^1$}
\email[]{hmdrpthk@gmail.com}
\author{Malaya K. Nayak$^2$}
\email[]{mk.nayak72@gmail.com}
\author{Nayana Vaval$^1$}
\author{Sourav Pal$^{3}$}
\affiliation{$^1$Electronic Structure Theory Group, Physical Chemistry Division, CSIR-National Chemical Laboratory, Pune, 411008, India}
\affiliation{$^2$Theoretical Chemistry Section, Chemistry Group, Bhabha Atomic Research Centre, Trombay, Mumbai 400085, India}
\affiliation{$^3$Department of Chemistry, Indian Institute of Technology Bombay, Powai, Mumbai 400076, India}
\begin{abstract}
The effective electric field experienced by the unpaired electron in the ground state of PbF, which is a potential candidate in the search 
of electron electric dipole moment due to some special characteristics, is calculated using Z-vector method 
in the coupled cluster single- and double- excitation approximation with four component Dirac spinor.
This is an important quantity to set the upper bound limit of the electron electric dipole moment.
Further, we have calculated molecular dipole moment and parallel magnetic hyperfine structure constant (A$_\|$) of $^{207}$Pb in PbF
to test the accuracy of the wave function obtained in the Z-vector method. 
The outcome of our calculations clearly suggests that the core electrons have significant contribution to the ``atom in
compound (AIC)'' properties.
\end{abstract}
\maketitle
\section{Introduction}
The well established model of the interaction of elementary particles, the Standard Model (SM), is incomplete as it cannot explain
some of the well known phenomena of fundamental physics. One such phenomenon is the dominance of matter over antimatter in our universe,
although the SM treats matter and antimatter exactly in the same way \cite{dine_2004}. The violation of two fundamental symmetries: inversion
symmetry (P) and charge conjugation (C), is one of the several conditions that can explain the matter antimatter asymmetry \cite{sakharov_1967}.
The CP violation within the SM originating from the complex quark mixing Kobayashi-Maskawa matrix is too weak to explain such an
asymmetry. Therefore, the search for an extra CP violation (flavour-diagonal CP violation), which is absent in the SM is needed to
explore new physics beyond the conventional SM \cite{pospelov_2005, fortson_2003}.
Another fundamental symmetry is the time-reversal (T) symmetry, which is also violated with the violation of CP symmetry though a
direct observation of the violation of T symmetry is yet to observe \cite{cpt_1964}. 

The electric dipole moment (EDM) of any elementary particle is a consequence of violation of both T and P as dipole
moment is odd under P and even under T but spin is even under P and odd under T \cite{bernreuther_1991}.
According to SM, the electron EDM
is too small (less than 10$^{-38}$) to observe experimentally \cite{khriplovich_2011}. Therefore, a measurable non zero EDM of electron would be
the proof of an extra CP violation and the first direct observation of T violation \cite{time_rev_1987}. However, the intensive search
for the electron EDM over the period of half a century have not drawn any conclusion to the final value of electron EDM, which would have out-turned
in the upper bound limit of the electron EDM for different quantum systems. Till date, the best limit of electron EDM in an atomic 
system is achieved from the Tl atom experiment ($|d_e| < 1.6 \times 10^{-27}$ e cm) \cite{tl_edm}.
However, the discovery of Sandars \cite{sandars_1967} reveals that the effective internal electric field experienced by an electron is profoundly enhanced
in heavy-atom containing polar molecule which makes these polar diatomics very promising candidate in the search of
the P- and P,T -violating experiments and creates a dimension to explore new physics. 
The latest best upper limit of electron EDM is set from the ThO experiment
($|d_e| < 8.7 \times 10^{-29}$ e cm) by ACME collaboration \cite{tho_edm}. This limit is one order lower in magnitude than the
previous best limit ($|d_e| < 10.5 \times 10^{-28}$ e cm), which is obtained from the molecular YbF experiment \cite{ybf_edm}.

Another advantage of using diatomic molecules in the search of electron EDM is the $\Omega$ doublet structure
of the ground and metastable state of such polar diatomic molecules, which gives an additional enhancement and due to this reason
different molecules having $^{2}\Sigma_{\frac{1}{2}}$ ground state (YbF \cite{ybf_edm, ybf_1, ybf_2, ybf_3, ybf_4},
HgF \cite{hgf_1} , HgH, BaF \cite{baf_1}), $^{2}\Pi_{\frac{1}{2}}$
ground state (PbF \cite{mcraven_2008, petrov_2013, skripnikov_2014}), $^{3}\Delta_{1}$ metastable state
(ThO \cite{tho_edm, tho_3, tho_4, tho_7}, ThF$^{+}$ \cite{thf_1, thf_2, thf_3},
HfH$^{+}$ \cite{hfh_pth}, PtH$^{+}$ \cite{hfh_pth}, WC \cite{wc_1, wc_2}) et cetera have been
proposed. Among these molecules, PbF has some interesting characteristics, which make it a strong
candidate in the search of electron EDM.
Pb is neither a lanthanide or actinide f element nor
a transition d element and PbF has a $^{2}\Pi_{\frac{1}{2}}$ ground state, which means
the unpaired electron is in the $\pi$ orbital while in most of the other molecules, the unpaired electron
is in the $\sigma$ orbital in their ground state.
PbF, being a $^{2}\Pi_{\frac{1}{2}}$ ground state molecule, the spin angular momentum of the unpaired electron contributing to the
magnetic moment almost cancels the orbital angular momentum contribution to the magnetic moment.
This leads to a smaller g-factor in the $^{2}\Pi_{\frac{1}{2}}$ state of PbF.
The smaller g-factor \cite{mawhorter_2011, skripnikov_2015} makes it
very insensitive to the background magnetic field and this leads to reduction in some systematic errors
in the experimental observation of electron EDM \cite{shafer-ray_2006}.
The other molecules having small g-factor like PbF are ThO, ThF$^{+}$, HfH$^{+}$, PtH$^{+}$, WC, et cetera but that is in their
metastable $^{3}\Delta_{1}$ state.
On the other hand, $^{2}\Pi_{\frac{1}{2}}$ state of PbF is a ground state, which is
easy to synthesize experimentally as compared to the metastable states.
The energy shifts of the levels of opposite parity in the ground
rotational state of $^{207}$PbF due to the $\Omega$ doubling is canceled by the magnetic hyperfine interaction
as a repercussion the gap between two opposite parity levels is very small (almost degenerate) \cite{mcraven_2008}.
Therefore, the molecule can be polarized very easily with the application of a weak electric field and
opposite sign of the $\Omega$ doublet component leads to the cancellation of some systematic error.

The effective electric field ($E_{eff}$) experienced by the electron in an atom or a molecule, which is equally known as P,T-odd
interaction constant (W$_d = E_{eff}$/$|\Omega|$) is a non-measurable quantity. On the other hand, it is very important to set
the upper bound limit in the search of electric dipole moment of electron. 
Therefore, one has to rely on a very accurate theoretical method to calculate $E_{eff}$,
precisely.
The calculation of E$_{eff}$ of a heavy diatomic
molecule is not a trivial task as it requires simultaneous inclusion of both the effect of special relativity and electron correlation
due to the intertwined nature of these two effects. The relativistic coupled cluster method using four-component wavefunction
meets these requirements to fulfill the purpose \cite{kaldor_1997}. 
There are two alternate choice to evaluate one electron response properties in the normal coupled cluster framework: 
(i) the expectation value approach and (ii) the derivative approach. The energy derivative approach is superior
to the other one as the calculated property value is very close to the property value calculated using full configuration-interaction (FCI) method.
The Z-vector method is a technique in the derivative framework, which simplifies the complexity associated with it.   
It calculates the energy derivate in a size extensive manner and capable of rendering accurate wave function 
both in the near nuclear region and apart from the nucleus and its performance has already been tested \cite{sasmal_2015}.

In this article, we have chosen Z-vector method in the coupled-cluster single- and double- excitation approximation (CCSD) for the
calculation of effective electric field, $E_{eff}$, experienced by the unpaired electron in the ground state of PbF molecule.
The parallel magnetic hyperfine structure (HFS) constant of $^{207}$Pb in PbF molecule is also calculated to judge the
accuracy in the calculated $E_{eff}$ values,
since both of these properties need an accurate wavefunction in the near nuclear region.
Further, we have calculated molecular dipole moment of PbF molecule and both the calculated HFS constant and molecular dipole moment
are compared with the experimental value
and all these results are compared with the values calculated by means of other theoretical methods.

The manuscript is organized as follows. A brief overview of the Z-vector method including concise details of E$_{eff}$ and
parallel component of the magnetic HFS constant are
described in Sec. \ref{sec2}. Computational details are given in Sec. \ref{sec3}. We presented our calculated
results and discuss about those in Sec. \ref{sec4} before making final remark in Sec. \ref{sec5}.
Atomic unit is used consistently unless stated.

\section{Theory}\label{sec2}
The theoretical estimation of $E_{eff}$ can be obtained by evaluating the following matrix element
\begin{eqnarray}
 E_{eff} = W_d|\Omega|  = \langle \Psi_{\Omega} | \sum_j^n \frac{H_d(j)}{d_e} | \Psi_{\Omega} \rangle ,
 \label{E_eff}
\end{eqnarray}
where $\Omega$ is the projection of total angular momentum along the molecular axis and $\Psi_{\Omega}$ is the
electronic wavefunction corresponding to $\Omega$ state. $n$ is the total number of electrons and 
H$_d$ is the interaction Hamiltonian of d$_e$ with internal electric field and is given by
\begin{eqnarray}
 H_d = -2icd_e \gamma^0 \gamma^5 {\bf \it p}^2 ,
 \label{H_d}
\end{eqnarray}
where $\gamma$ are the usual Dirac matrices and {\bf \it p} is the momentum operator.

The parallel magnetic HFS constant in Dirac theory can be derived by taking the z projection (along molecular axis)
of the expectation value of the corresponding Hamiltonian and is given by
\begin{eqnarray}
 A_{\|}= \frac{\vec{\mu}_k}{I\Omega} \cdot \langle \Psi_{\Omega} | \sum_i^n
\left( \frac{\vec{\alpha}_i \times \vec{r}_i}{r_i^3} \right)_{z} | \Psi_{\Omega}  \rangle,
 \label{A_para}
\end{eqnarray}
where $I$ is the nuclear spin quantum number, $\alpha$ are the usual Dirac matrices and $\vec{\mu}_k$ is the magnetic
moment of the nucleus $k$.

The dynamic part of the electron correlation effect is included by the exponential structure of the coupled cluster
wavefunction, which is given as
\begin{eqnarray}
 | \Psi_{cc} \rangle = e^{T} | \Phi_0 \rangle ,
 \label{cc_eqn}
\end{eqnarray}
where, $|\Phi_0\rangle$ is the ground state single determinant wavefunction and $T$ is the excitation operator.
The form of the $T$ operator is given by
\begin{eqnarray}
 T=T_1+T_2+...+T_N=\sum_n^N T_n,
\end{eqnarray}
where
\begin{eqnarray}
T_m= \sum\limits_{\stackrel{i<j\dots}{a<b\dots}}
t_{ij \dots}^{ab\dots}{a_{a}^{\dagger}a_{b}^{\dagger} \dots a_{j} a_{i}} ,
\end{eqnarray}
where i,j(a,b) are hole(particle) index and t$_{ij..}^{ab..}$ are the cluster amplitudes corresponding 
to the cluster operator $T_m$.
The equations of cluster amplitude can be obtained by pre-projecting excited determinant with respect to $|\Phi_0\rangle$
of the above equation. In coupled cluster single and double model, T is T$_1$ + T$_2$. The equations for T$_1$ and T$_2$ are
given as
\begin{eqnarray}
 \langle \Phi_{i}^{a} | (H_Ne^T)_c | \Phi_0 \rangle = 0 , \,\,
  \langle \Phi_{ij}^{ab} | (H_Ne^T)_c | \Phi_0 \rangle = 0 ,
 \label{cc_amplitudes}
\end{eqnarray}
where $|\Phi_{i}^{a}\rangle$ and  $|\Phi_{ij}^{ab}\rangle$ are singly and doubly excited determinant, respectively and
H$_N$ is the normal ordered Dirac-Coulomb Hamiltonian. The subscript $c$ means only the connected terms exist in the
contraction between H$_N$ and T, which ensures the size-extensivity.
As, the normal coupled cluster (NCC) is not a 
variational method, the energy is not optimized with respect to the determinantal coefficients in the expansion of the
many electron correlated wavefunction and the molecular orbital (MO) coefficients for a fixed nuclear geometry. Thus the
calculation of energy derivative in NCC framework needs to include the derivative of energy with respect to determinantal
coefficients and MO coefficients in addition to the derivative of these two coefficients with respect to the external
perturbation field. Thus one needs to calculate these terms for each external field of perturbation. However, in
Z-vector method, this can be avoided with the introduction of a de-excitation operator, $\Lambda$, where the $\Lambda$
amplitude equations are linear and perturbation independent.
The second quantized form of the $\Lambda$ operator is given by
\begin{eqnarray}
 \Lambda=\Lambda_1+\Lambda_2+...+\Lambda_N=\sum_n^N \Lambda_n,
\end{eqnarray}
where
\begin{eqnarray}
 \Lambda_m= \sum\limits_{\stackrel{i<j\dots}{a<b\dots}}
\lambda_{ab \dots}^{ij\dots}{a_{i}^{\dagger}a_{j}^{\dagger} \dots a_{b} a_{a}} ,
\end{eqnarray}
where i,j(a,b) are the hole(particle) indices and $\lambda_{ab..}^{ij..}$ are the cluster amplitudes corresponding 
to the cluster operator $\Lambda_m$.
The detailed description of $\Lambda$ operator and
$\Lambda$ amplitude equation can be found in Ref. \cite{bartlett_zvec}. In CCSD model, $\Lambda$ becomes, $\Lambda=\Lambda_1+\Lambda_2$.
The explicit equations for the amplitudes of $\Lambda_1$ and $\Lambda_2$ operators are
 \begin{eqnarray}
  \langle \Phi_0 |& [\Lambda (H_Ne^T)_c]_c | \Phi_{i}^{a} \rangle + \langle \Phi_0 | (H_Ne^T)_c | \Phi_{i}^{a} \rangle = 0,
  \label{lambda_1}
\end{eqnarray}
\begin{eqnarray}
  \langle \Phi_0 |& [\Lambda (H_Ne^T)_c]_c | \Phi_{ij}^{ab} \rangle + \langle \Phi_0 | (H_Ne^T)_c | \Phi_{i}^{a} \rangle \nonumber\\
 &  \langle \Phi_{i}^{a} | \Lambda | \Phi_{ij}^{ab} \rangle + \langle \Phi_0 | (H_Ne^T)_c | \Phi_{ij}^{ab} \rangle = 0.
\label{lambda_2}
\end{eqnarray}
Although the term $\langle \Phi_0 | (H_Ne^T)_c | \Phi_{i}^{a} \rangle \langle \Phi_{i}^{a} | \Lambda | \Phi_{ij}^{ab} \rangle$
in $\Lambda_2$ equation (equation \ref{lambda_2}) results into one disconnected diagram but the diagram is not of the kind of closed
with disconnected part, it is linked (for details see Ref. \cite{sasmal_2015}). This ensures the extensivity.
The equation for energy derivative can be written as
\begin{eqnarray}
\Delta E' = \langle \Phi_0 | (O_Ne^T)_c | \Phi_0 \rangle + \langle \Phi_0 | [\Lambda (O_Ne^T)_c]_c | \Phi_0 \rangle,
\end{eqnarray}
where $O_N$ is the normal ordered property operator.
\section{Computational Details}\label{sec3}
The DIRAC10 \cite{dirac10} program package is used to construct one electron spinors, two-body matrix elements and one-electron property integrals.
Gaussian charge distribution is considered to take care of the finite size of the nucleus where the nuclear parameters \cite{visscher_1997} are taken as
default value of DIRAC10. Restricted kinetic balance (RKB) \cite{dyall_2006} is used to construct small component basis functions from large component
basis. In RKB, the basis functions are represented in scalar basis and unphysical solutions are removed by diagonalizing free particle
Hamiltonian. The positive and negative energy solutions are generated in 1:1 manner by this formalism. We have done five different
calculations (A-E) by varying basis function and number of correlated electrons. For Pb, dyall.cv3z \cite{pb_basis}
and for F, cc-pCVTZ \cite{f_basis} basis is used
and two different calculations are done by using 55 and 73 number of correlated electrons and these are denoted by A and B, respectively.
We have done three more calculations by using 55, 73 and 91 (all electron) correlated electrons where dyall.cv4z \cite{pb_basis} and
cc-pCVQZ \cite{f_basis} are
used for Pb and F, respectively and these calculations are denoted as C, D and E, respectively. The cutoff used for A, B, C, D and E
calculations are 3500 a.u., 1000 a.u., 70 a.u., 70 a.u., and 70 a.u., respectively. We have used the experimental bond length
(3.89 a.u.) \cite{bond_len}
for the calculation of properties of PbF in its ground state.
\begin{table}[ht]
\caption{Dipole moment, parallel magnetic HFS of $^{207}$Pb and effective electric field of PbF }
\begin{ruledtabular}
\newcommand{\mc}[3]{\multicolumn{#1}{#2}{#3}}
\begin{center}
\begin{tabular}{lrrrrr}
 & \mc{2}{c}{$\mu$ (D)} & \mc{2}{c}{A$_{\|}$ (MHz)} & $E_{eff}$(GV/cm)\\
\cline{2-3} \cline{4-5}\\
Basis  & Z-vector & Expt. \cite{mawhorter_2011} & Z-vector & Expt. \cite{mawhorter_2011, petrov_2013}& Z-vector\\
\hline
A & 3.71 & & 9865 &   & 36.6\\
B & 3.72 & & 9962 &   & 37.5\\
C & 3.82 & 3.5$\pm$0.3 & 9968 & 10147 & 37.2\\
D & 3.82 & & 10043 &   & 37.9\\
E & 3.83 & & 10121 & & 38.1
\end{tabular}
\end{center}
\end{ruledtabular}
\label{pbf}
\end{table}
\begin{figure}[b]
\centering
\begin{center}
\includegraphics[height=4.5 cm, width=7.5 cm]{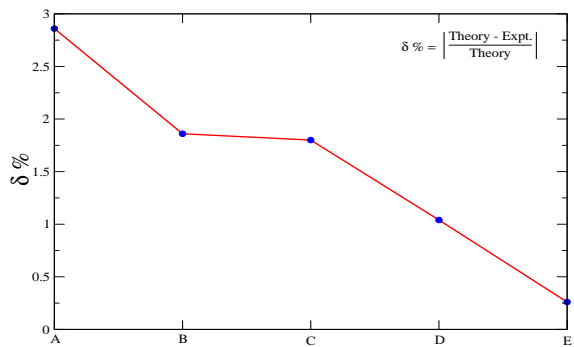}
\caption {Relative deviations between Z-vector and Expt. values of parallel magnetic HFS values.}
\label{figure}
\end{center}  
\end{figure}
\section{Results and Discussions}\label{sec4}
The properties described by Eqs. \ref{E_eff}, \ref{A_para} are also known as ``atom in compound (AIC)''
properties as these properties strongly depend on the electronic configuration of the given (heavy) atom
instead of the bonds between atoms \cite{titov_2014}.
In Table \ref{pbf}, we present the molecular dipole moment ($\mu$), parallel component of magnetic hyperfine structure
constant (A$_{\|}$) of $^{207}$Pb and effective electric field experienced by the unpaired electron of PbF.
From Table \ref{pbf}, it is clear that our dipole moment values are in good agreement with the experimental
value \cite{mawhorter_2011}. We got values in the range from 3.71 D (basis A) to 3.83 D (basis E) due to different basis and number of
correlated electrons but this range fits well within experimental limit (3.5$\pm$0.3 D).

The calculated parallel
component of magnetic HFS constant of $^{207}$Pb in PbF shows an excellent agreement with the experiment \cite{mawhorter_2011, petrov_2013}; specially
for E basis where the absolute difference between theory and experiment is only 26 MHz. The relative error of the 
parallel magnetic HFS constant in five different calculations (A-E) is shown in Fig. \ref{figure}. The highest and lowest deviations
of Z-vector value from experiment are for basis A (2.86\%) and basis E (0.26\%), respectively. This trend in the deviations
(expressed in $\delta$\%) is expected. When we go from triple zeta (TZ) basis to quadruple zeta (QZ) basis with same number
of correlated electrons (from A to C and B to D where number of correlated electrons are 55 and 73, respectively) the
$\delta$\% decreases as QZ improves the configuration space more by including one higher angular momentum basis function than TZ.
On the other hand, for same basis, if we include more electrons in correlation calculation
(for TZ, from A to B and for QZ, from C to E), the $\delta$\% decreases
as the more number of correlated electrons includes more orthogonal space to Dirac-Hartree-Fock space and thus includes more
correlation contribution to the property value. It is also interesting to see that in TZ basis, when we go from A to B,
the addition of 18 electrons (i.e., 4s+3d+4p core electrons of Pb) improves the parallel magnetic HFS constant by 97 MHz.
In QZ basis, as we go from C to D and D to E,
the addition of 18 electrons (i.e., 4s+3d+4p and 1s-3p core electrons of Pb, respectively) improves the A$_{\|}$ value by 75 MHz
and 78 MHz, respectively.
From this observation we can conclude that
the core electrons have significant role in the correlation contribution of parallel magnetic HFS value.

In Table \ref{pbf}, we present our Z-vector results of $E_{eff}$ of PbF in five different calculations where the value
ranges from 36.6 GV/cm to 38.1 GV/cm. We believe that the value in E basis (38.1 GV/cm) is the most reliable value of
$E_{eff}$ of PbF system as its corresponding parallel magnetic HFS value has the closest agreement with experiment.
In E basis, the Z-vector magnetic HFS value has a uncertainty of 0.26\%. So, considering basis set and other higher order
correlation and relativistic effects, we can conclude that the $E_{eff}$ of PbF is 38.1 GV/cm with 4\% uncertainty.
\begin{table}[b]
\caption{ Comparison of molecular dipole moment, magnetic HFS constant and $E_{eff}$ of PbF }
\begin{ruledtabular}
\newcommand{\mc}[3]{\multicolumn{#1}{#2}{#3}}
\begin{center}
\begin{tabular}{lrrr}
Method & \mc{1}{c}{$\mu$} & \mc{1}{c}{A$_{\|}$ ($^{207}$Pb)} & $E_{eff}$\\
× & (Debye) & (MHz) & (GV/cm) \\
\hline
SODCI(13e) \cite{baklanov_2010} & 4.26 & 9727 & 33 \\
SODCI(13e)+OC \cite{baklanov_thesis} & 5.00 & 10262 & 37\\
2c-CCSD(31e) \cite{skripnikov_2014} & 3.97 & 10265 & 41\\
2c-CCSD(T)(31e) \cite{skripnikov_2014} & 3.87 & 9942 & 40 \\
4c-Z-vector(QZ, all electron) & 3.83 & 10121 & 38.1 \\
Experiment \cite{mawhorter_2011, petrov_2013} & 3.5 $\pm$ 0.3 & 10147 & ×
\end{tabular}
\end{center}
\end{ruledtabular}
\label{comparison}
\end{table}

We compared our Z-vector result with other theoretically obtained values. From Table \ref{comparison}, it is clear
that our all electron value in QZ basis for both dipole moment and parallel magnetic HFS of $^{207}$Pb in PbF
has the best agreement with experiment among all the other theoretical values. Baklanov {\it et al} did two 
calculations with 13 correlated electrons in spin-orbit direct CI (SODCI) methods -- one without outer
core (OC) \cite{baklanov_2010} correlation correction and the other with OC correlation correction \cite{baklanov_thesis}.
The SODCI with OC correction \cite{baklanov_thesis} calculation
gives better value for A$_\|$ of $^{207}$Pb but gives poorer value of molecular dipole moment. It is worth to remember
that CI is not size extensive and thus does not scale properly with number of electrons. So, CI is not a reliable
method for the system with a reasonable number of electrons, especially with heavy atom containing systems.
Recently, Skripnikov {\it et al} \cite{skripnikov_2014} have done two two-component (2c) coupled cluster calculations -- one with
single and double approximation (CCSD) and the other with CCSD with perturbative triples (CCSD(T)) correction.
In their calculations, Skripnikov {\it et al} have included only 31 correlated electrons and removed
60 electrons (1s-4f inner-core electrons of Pb) by using ``valence'' semilocal version of the GRECP
scheme \cite{mosyagin_2010, titov_1999}. In valence semilocal version of GRECP, the components
were constructed for nodal valence pseudospinors by interpolating the potential in the neighbor of pseudospinor node
to avoid the singularity in the potential. 
The problem of valence GRECP approximation is that it can lead to ``non-negligible''
errors for valence electronic states due to the improper reproduction of nuclear screening \cite{mosyagin_2010}.
Although the molecular GRECP calculations are two-component ones, the proper four-component wave function near the nucleus
is restored at the nonvariational restoration stage that can lead to small errors. On the other hand our all-electron
calculations are four-component at all the stages of calculations.
Although the authors in Ref. \cite{skripnikov_2014} claim that the 
``contemporary full-electron studies have not yet been able to unambiguously surpass our approach when it comes to
AIC and spectroscopic properties of interest'', we believe that the explicit treatment of core electrons is
necessary for some AIC properties where the polarization of the inner core electrons plays an important role,
which is evident from our calculated parallel magnetic HFS constant value.
\section{Conclusions}\label{sec5}
In conclusion, we have applied Z-vector method in the coupled cluster framework to calculate $E_{eff}$
experienced by the electron in the ground state of PbF molecule.
The calculated molecular dipole moment and A$_\|$ of $^{207}$Pb are in excellent agreement with the experimental values.
As the calculated HFS constant is in very good agreement with experiment, we can say that our calculated $E_{eff}$ = 38.1 GV/cm
is most reliable
as both require accurate wave function near the nucleus and expectation value of their operator are similar in structure.
The core electrons have significant contribution in the calculated values, which is evident from our calculated results. 
Therefore, it desirable to treat all the electrons explicitly to have much more accurate and reliable result. 
%
\section*{Acknowledgement}
Authors acknowledge a grant from CSIR XII$^{th}$ Five Year
Plan project on Multi-Scale Simulations of Material (MSM) and the resources
of the Center of Excellence in Scientific Computing at CSIR-NCL.
S.S. and H.P acknowledge the CSIR for their fellowship.
We thank Dr. L. V. Skripnikov for his valuable comments and suggestion.

\end{document}